\documentstyle[aps,12pt]{revtex}
\input{epsf}

\newfont{\largemi}{cmmi10}

\newfont{\smallmi}{cmmi6}

\begin{document}
\draft

\title{Generic Rotation in  a Collective $SD$ Nucleon-Pair Subspace  }
\author{ Y. M. Zhao$^{a,b,c}$, S. Pittel$^{a}$,
R. Bijker$^{d}$\footnote{On sabbatical leave at
Dipartimento di Fisica, Universita degli
Studi di Genova, Via Dodecaneso 33,
I-16146 Genova, Italy}, A. Frank$^{d,e}$, and   A. Arima$^{f}$ }

\vspace{0.2in}
 \address{
$^a$ Bartol Research Institute, University of Delaware,
Newark, DE19716-4793, USA  \\
$^b$ Department of Physics, Saitama  University,
Urawa-shi, Saitama 338-0825, Japan \\
$^c$ Department of Physics, Southeast University,
Nanjing, 210018, P.R. China \\
$^d$ Insituto de Ciencias Nucleares,
Universidad Nacional Aut\'onoma de M\'exico, \\
Apartado Postal 70-543, 04510 M\'exico, D. F., M\'exico \\
$^e$Centro de Ciencias F{\'{\i}}sicas,
Universidad Nacional Aut\'onoma de M\'exico, \\
Apartado Postal 139-B, 62251 Cuernavaca, Morelos, M\'exico \\
$^f$ The House of Councilors, 2-1-1 Nagatacho,
Chiyodaku, Tokyo 100-8962, Japan }

\vspace{0.2in}

\date{\today}
\maketitle

\begin{abstract}

Low-lying collective states involving many nucleons interacting by
a random ensemble of two-body interactions (TBRE) are investigated
in  a collective  $SD$-pair subspace, with the collective pairs
defined dynamically from the two-nucleon system. It is found that
in this truncated pair subspace collective vibrations arise
naturally for a general TBRE hamiltonian whereas collective
rotations do not. A hamiltonian restricted to include only a few
randomly generated separable terms is able to produce collective
rotational behavior, as long as it includes a reasonably strong
quadrupole-quadrupole component. Similar results arise in the full
shell model space. These results suggest that the structure of the
hamiltonian is key to producing generic collective rotation.

\end{abstract}
\vspace{0.1in}


\vspace{0.4in}

\newpage

Recently Johnson and collaborators studied the low-lying spectra
of many-body systems in the presence of random interactions.
Surprisingly, they found that patterns of more orderly
interactions persist, such as a predominance of total angular
momentum $I^{\pi}=0^+$ ground states with large gaps relative to
excited states [1-3]. The subject of many-body systems interacting
by a two-body random ensemble (TBRE) has been attracting much
interest since this discovery.  Many authors have tried  to
understand the regularities exhibited by a many-body system
interacting randomly and
to uncover other robust properties of many-body systems [4-22].

In Ref. \cite{Bijker} it was shown that  both vibrational and
rotational features arise in the interacting boson model (the IBM)
\cite{Iachello} in the presence of random interactions. In
contrast, as shown in many works,  rotational behavior does not
generically arise in fermion systems interacting via random
interactions. It was suggested, therefore, that a special feature
of the ensemble might be necessary to obtain a generic rotational
behavior in fermion systems \cite{Johnson3}.

The remarkable rotational peak obtained in  Ref. \cite{Bijker} is
suggestive that a key to obtaining a rotational peak from a random
hamiltonian might be to restrict the space to a collective
subspace built up from the lowest $S$ and $D$ pairs, since these
are the objects that are represented in the IBM by its $s$ and $d$
bosons \cite{Iachello}. To see  whether this is  true, we report
here calculations in which random interactions are used in a
truncated version of the shell model in which only collective $S$
and $D$ pairs are retained. We limit ourselves, however, to
systems of identical nucleons

Another possibility is that generic rotation requires the
hamiltonian to have some specific features and is not just a
consequence of the space. A similar philosophy was recently
discussed by Vel\'azquez and Zuker \cite{Zuker}, who showed that a
displaced random hamiltonian which is attractive on average
usually leads to rotational motion. Here we consider an
alternative way of modifying the hamiltonian -- as a limited sum
of separable interactions with random strengths -- in order to
study under which conditions it is possible to generate rotational
features in the calculated spectra.

This can be readily implemented within the
context of a truncated $SD$-pair space.
A natural way to carry out such calculations is through the use of
the nucleon pair approximation (NPA) \cite{Chen,unified}. The NPA
is very similar in spirit to the generalized seniority scheme
\cite{Talmi} and the broken pair approximation \cite{Gambhir},
which were  used extensively in efforts to establish
a microscopic foundation for the IBM \cite{Otsuka}. The NPA, however,
has several computational advantages over these earlier formulations,
and is therefore the method we will employ in this work.

The NPA calculations we report here were carried out using the
following strategy. First, we selected only those sets of random
interactions/parameters for which the hamiltonian produced an
$I^{\pi}=0^+$ ground state and an $I^{\pi}=2^+$ first excited
state, when diagonalized for a system of two identical nucleons,
and then assumed that the corresponding wave functions represent
our collective $S$ and $D$ pairs, respectively. The $S$ and $D$
pairs obtained in such a way, though not fully self-consistent,
should be good enough to permit a meaningful study of the
regularities of the $SD$-pair approximation in the presence of
random interactions. For each such case, the same two-body
interaction was then used to calculate the spectrum for six
identical nucleons, with the procedure iterated until there are
acceptable statistics.  In the analysis, we focused only on those
cases for which the six particle system had a $0^+$ ground state,
calculating for them the ratio $R=(E_{4_1^+}-E_{0_1^+}) /
(E_{2_1^+} - E_{0_1^+}) $. For vibrational systems,  $R= 2$, and
for rotational systems $R=10/3$. Rotational motion will be said to
be generic for an $SD$-pair system if a clear peak with $R$ around
$10/3=3.33$ appears.

It is useful at this point to briefly review the NPA formalism
that has been used in these calculations.  It begins with the
introduction of an operator $A^{(r)\dagger}_{\mu\phantom{\dagger}}$
which creates a collective pair of angular momentum $r$ and
z-projection $\mu$ and which is defined by
\begin{equation}
  A^{(r) \dagger}_{\mu\phantom{\dagger}}
               =   \sum_{ab} y(abr)
   \left( C^{\dagger}_{a}\times C^{\dagger}_{b} \right)^{(r)}_{\mu}
   ~.
\end{equation}
Here $y(abr)$ are structure coefficients of the collective pair,
and $r$=0 and 2 correspond to the $S$ or $D$ pairs, respectively.
The $C_a^{\dagger}$  and $C_b^{\dagger}$ are single-particle
creation operators with $a, b$ denoting the respective
single-particle orbit, including its  $j$ value.
  These   pairs are coupled
 stepwise  to yield an $N$-pair basis
$|\tau J_N \rangle = A ^{(J_N)\dagger}(r_i,J_i) |0 \rangle$ with
\begin{equation}
A^{(J_N)\dagger}(r_1r_2...r_N, J_1J_2...J_N)  =
\left(...( A^{(r_1)\dagger} \times A^{(r_2) \dagger} )^{(J_2)}
\times ... \times A^{(r_N)\dagger} \right)^{(J_N)},
\end{equation}
where  $J_1= r_1$,  and $J_N$ is the total angular momentum of the above
$N$-pair operator.

The restricted separable hamiltonians that we  consider in
this work can be written in the form
\begin{equation}
H=H_0 + H_{\rm P} +   V_{\rm ph} ~,     \label{sep1}
\end{equation}
where $H_0$, $H_{\rm P}$ and $V_{\rm ph}$ are
the spherical single-particle energy term,
 generalized pairing,  and
particle-hole type interactions, respectively.
The generalized pairing interaction, $H_{\rm P}$,  is defined as
\begin{equation}
H_{\rm P} = V_0 + V_2 + \cdots, \label{sep2}
\end{equation}
where $V_0$ is the monopole pairing interaction, defined as
\begin{equation}
 V_0=  G_0  {\cal P}^{(0)\dagger}  {\cal P}^{(0)},   ~~
{\cal P}^{(0)\dagger}_{} =   \sum_{a_{}}
\frac{\hat{j}_{}}{2}(C_{a_{}}^{\dagger} \times
C_{a_{}}^{\dagger})_0^{(0)} ~,
\end{equation}
with  $\hat{j} =(2j+1)^{\frac{1}{2}}$, and $V_2$ is the quadrupole
pairing force, defined as
\begin{equation}
V_2 =   G_2{\cal P}^{(2) \dagger }_{} \cdot {\cal P}^{(2)},   ~~
{\cal P}^{(2)\dagger} = \sum_{a_{} b_{}} q(a_{} b_{}) \left(
C^{\dagger}_{a_{}} \times C^{\dagger}_{b_{}} \right)^{(2)} ~.
\label{sep3}
\end{equation}
The quantity $q(ab)$ appearing in the quadrupole pairing force is
precisely the same quantity that appears in the quadrupole
operator, namely
\begin{equation}
Q_{M} = \sum_{a b} q(a b) \left( C^{\dagger}_{a} \times
\tilde{C}_{b} \right)^{(2)}_M = \sum_{ambm'} \langle am
 |{\cal Q}^{(2)}_{M}| bm' \rangle
 C^{\dagger}_{am } C_{bm'},
 \label{quad}
\end{equation}
where   ${\cal Q}_M = r^2 {\rm Y}_{2M}$, and  $q(a b) =
\frac{(-)^{j+1/2}}{ \sqrt{20\pi} }
 \hat{j} \hat{j'} C_{j1/2, j' -1/2}^{20} {\cal R}$.
 $C_{j 1/2, j' -1/2}^{2 0} $ is a
Clebsch-Gordan coefficient, and
$ {\cal R} =    \langle n l |r^{2} |n l'   \rangle$ \cite{Yoshida}.
The particle-hole interaction takes the form
\begin{equation}
 V_{\rm ph} =\kappa Q  \cdot Q   +   \cdots.
 \label{sep4}
\end{equation}
The quadrupole-quadrupole piece is expressed in terms of the same
quadrupole operator given above (\ref{quad}).

A general nuclear hamiltonian for one kind of particle can be
written as
\begin{eqnarray}
H=H_0 - \frac{1}{4}  \sum_{abcd,J}  &&
\sqrt{(1+\delta_{ac})(1+\delta_{bd})} ~\hat{J} ~ \langle ab,J | V
| cd,J \rangle   \nonumber \\  &&\times \left[
\left(C_a^{\dagger}C_b^{\dagger}\right)^{(J)} \times \left(
C_c^{\dagger}C_d^{\dagger} \right)^{(J)} \right]^{(0)}. \label{V}
\end{eqnarray}
The various separable interactions can be readily transformed to
the general form of a two-body interaction given in (\ref{V}).
The procedure for calculating matrix elements of one- and two-body
interactions in the NPA is to first rewrite the matrix elements in
terms of $N$-pair overlaps $\langle \tau' J_N | \tau J_N \rangle$,
and then to calculate these $N$-pair overlaps in terms of
$N-1$-pair overlaps by using the Wick theorem for coupled clusters
developed in \cite{Wick}. Thus, the hamiltonian matrix elements are
calculated in a recursive way in the NPA.

We now return to the problem motivating this study -- to search
for generic rotation in a collective $SD$ nucleon-pair subspace.
First we take a general two-body hamiltonian. The two-body matrix
elements we use are defined as in Eq. (\ref{V}) and the
single-particle energies, $H_0$, are set to be zero.

In Fig. 1 we plot the distribution of $R$ values for six identical
nucleons in the $sd$, $pf$ and $sdg$ shells, respectively. We
first note that the distribution of $R$ values in the $sd$ shell
within an $SD$ subspace is similar to that obtained in the full
shell model space \cite{Johnson3} -- a broad distribution
extending to $R \sim 1.3$. When one goes to larger shells, the
distributions become sharper, and shift to the right from the $sd$
shell ($R \sim 1.3$) to the $sdg$ shell ($R\sim 1.91$).
Nevertheless, no sharp peak at $R\sim 3.33$ appears, even though
the  distribution does extend to that region. From this we
conclude that generic rotational collectivity in the shell model
does not seem to emerge from pair truncation of the space alone.
Furthermore, statistically the full shell model space and the $SD$
truncated subspace defined here give essentially the same results
for a general two-body interaction.

Since rotations do not seem to arise in a collective
$SD$-pair subspace if a TBRE hamiltonian is assumed, we turn now
to the second possibility, that generic rotation might appear if
we use a more-restrictive hamiltonian. As suggested earlier, we
will consider the possibility of using pairing plus particle-hole
type interactions, with their strength parameters generated
randomly.

In Figure 2, we show results for six identical nucleons in the
$sd$ shell based on a sum of three terms, monopole pairing,
quadrupole pairing and quadrupole-quadrupole. All are defined
according to (\ref{sep1}-\ref{sep4}). When all three interaction
strengths are treated on the same footing -- except that  $G_2$
and $\kappa$ are in units of $MeV/fm^4$, whereas $G_0$ is in units
of $MeV$ -- we arrive at the distribution of $R$ values shown in
Fig. 2a. In this case, no sharp rotational peak is observed.
Instead, a peak appears around $R\sim 1.3$, with a long tail
extending to $R\sim 3.1$.  If we artificially enhance the $Q \cdot
Q$ strength parameter $\kappa$ by a factor $\epsilon$, we arrive
at the results shown in Figs. 2b-d. As $\epsilon$ is increased,
i.e. as the quadrupole-quadrupole strength is enhanced, a peak at
$R\sim 3.1$ gradually appears. On the other hand, the probability
of $R>3.1$ remains very small.

As we progressively increase the size of the shell, the peak at
$R\sim 1.3 $ gradually disappears and another peak, very close to
$R\sim 3.3$, emerges. This is illustrated in Fig. 3 where we show
results for six identical nucleons in the $pf$, $sdg$, $pfh$, and
$sdgi$ shells with $\epsilon=1.0$. For a large shell, the peak at
$R \sim 3.3$ becomes very well pronounced.  When we examine the
calculated results more carefully, we find that when $R  \sim 3.33
$,   the ratio of $(E_{6_1^+}-E_{0_1^+}) / (E_{2_1^+} - E_{0_1^+})
$ is also very compact and close to $7$, the value in the
rotational limit.

It is interesting to see what happens if we use the same
restricted phenomenological hamiltonian in the full shell model
space. This can be readily done for the $sd$ shell, for which we
show the corresponding full shell model results in Fig. 4. In Fig.
4a, we limit ourselves to monopole pairing and a $Q\cdot Q$
interaction, while in Fig. 4b we include quadrupole pairing as
well.  One gets predominantly a rotational  peak with $R\sim
10/3$. When $\epsilon = 10$, we obtain a rotational peak at $R
\sim 10/3$, but this peak disappears as  we gradually reduce to
$\epsilon = 1$.

Based both on the pair-truncated results -- which we were able to
perform for many different shells -- and the full shell-model
results -- limited to the $sd$ shell -- we conclude that rotational
motion is related closely to the form of the two-body interaction.
In particular, for systems of identical nucleons there must be a strong
quadrupole-quadrupole component in the interaction for a
rotational peak to emerge.

To check the role of higher multipole interactions in producing
the $R\sim 3.3$ peak, in Fig. 5 we plot the distribution of $R$
values of  six identical nucleons in the $pfh$ shell interacting
by  a monopole pairing, quadrupole pairing, quadrupole-quadrupole
force, hexadecapole pairing  and hexadecapole-hexadecapole force.
The hexadecapole interaction that we use has no $r$ dependence
(i.e., the radial matrix element is assumed to be ${\cal R}=1$),
so we introduce a renormalization factor, 40, as `compensation'.
To isolate the role of the hexadecapole interactions, we introduce
an additional multiplicative factor $\epsilon$ for the
hexadecapole interactions. We note that the hexadecapole forces
suppress the $R \sim 3.3$ peak.  If one artificially enlarges the
hexadecapole forces, the $R \sim 3.3$ peak is further quenched.

Summarizing, we have analyzed in this paper the conditions for the
appearance of rotational motion for random hamiltonians in the context
of a system of identical nucleons restricted to collective $S$ and $D$
pairs. We find that in such a truncated pair subspace vibrations
arise generically for a general TBRE hamiltonian but rotations do
not. With appropriate restrictions on the form of the hamiltonian, we are
able to generate collective rotations, as had been found earlier for $sd$
boson systems \cite{Bijker}. In Ref. \cite{Bijker}, a TBRE
ensemble with 7 or 8 independent parameters was used in a space of
$s$ and $d$ bosons, while here we take a TBRE hamiltonian with
only 3 parameters for nucleons in many-$j$ shells. Not
surprisingly, the quadrupole-quadrupole interaction seems to play
a key role in obtaining a peak at $R\sim 3.33$. It was found, that all
other interaction terms tend to wipe out the rotational peak.
It is also noted that the results are not effected appreciably by the
inclusion of single-particle splittings.

It is interesting to ask why the interacting boson model with
fully random boson parameters is able to give rise to rotations,
while the shell model truncated to $SD$ pairs cannot. The answer
may lie in the fact
that the interacting boson model should not be thought of as
simply a pair truncation of the model space, but rather as a
truncation of the model space that arises from {\em the dominance
of quadrupole correlations}. Thus, the interacting boson model,
whether modelled by a random hamiltonian or not, is already a
consequence of quadrupole and pairing correlations. There is no
inconsistency, therefore, between the results of \cite{Bijker} on
the interacting boson model and those reported here.
It would be interesting to see whether these conclusions hold up
in the presence of both protons and neutrons.

One of us (YMZ) would like to acknowledge the warm hospitality of
the Bartol Research Institute, where most of this work was done,
and Dr. N. Yoshinaga for valuable discussions. This work was
supported in part by the Japanese Society for the Promotion of
Science under Contract No. P01021,  the  U.S.
National Science Foundation under Grant No. PHY-9970749,
by CONACyT under project Nos. 32416-E and 32397-E,
and by DGAPA under project No. IN106400.

\newpage

\vspace{0.3in}
{\bf captions}:

FIG. 1 ~ The distribution of   $R$ values for six identical
nucleons interacting by a general TBRE hamiltonian in a) the $sd$
shell, ~ b) the $pf$ shell, and c) the $sdg$ shell.

\vspace{0.3in}

FIG. 2 ~ The distribution of  $R$ values for six identical nucleons in the
$sd$ shell interacting by a random monopole pairing, quadrupole
pairing, and quadrupole-quadrupole force.  The strength $\kappa$
of the quadrupole-quadrupole interaction is multiplied by a factor
$\epsilon$ to assess the importance of the quadrupole-quadrupole
interaction.

\vspace{0.3in}

FIG. 3 ~ The distribution of  $R$ values for six identical nucleons
interacting by a random monopole pairing, quadrupole pairing, and
quadrupole-quadrupole force for a)  the $pf$ shell, b)  the $sdg$
shell, c)  the $pfh$ shell, d)  the $sdgi$ shell. In all cases,
$\epsilon=1.0$.

\vspace{0.3in}

FIG. 4 ~  The distribution of  $R$ values for six identical nucleons in a
$sd$ shell interacting by a) monopole pairing plus a
quadrupole-quadrupole force, and  b)  monopole and quadrupole
pairing plus a quadrupole-quadrupole force.
The c) and d) are the same as a) and b), respectively,  except that
a quadrupole enhancement of $\epsilon=10$ was used.

\vspace{0.3in}

FIG. 5 ~ The distribution of $R$ of six identical nucleons in the
$pfh$ shell interacting by  a monopole pairing, quadrupole pairing,
quadrupole-quadrupole force, hexadecapole pairing  and
hexadecapole-hexadecapole force. The  hexadecapole interactions
have no $r$ dependence, so  we use 40 as a
renormalization factor.  To see the effect of
 hexadecapole interactions, we
multiply an adjustable factor, $\epsilon$, on them.
a) $\epsilon=1$, and b) $\epsilon=25$.

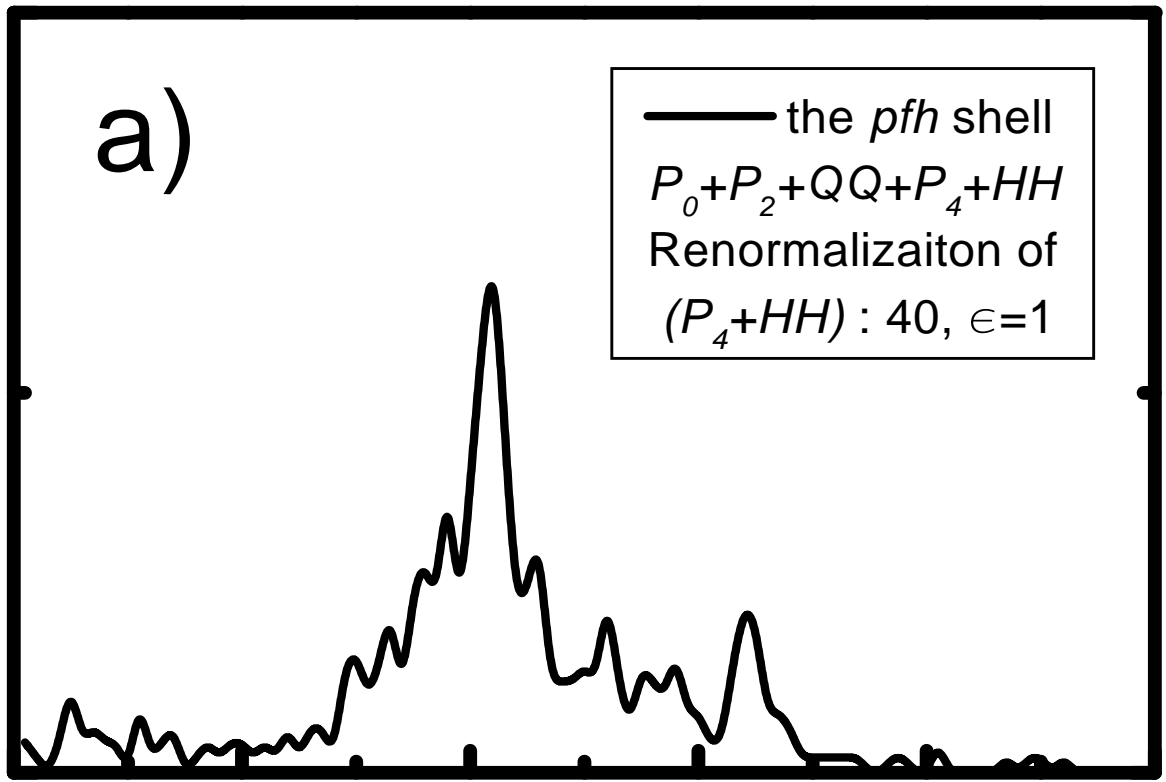
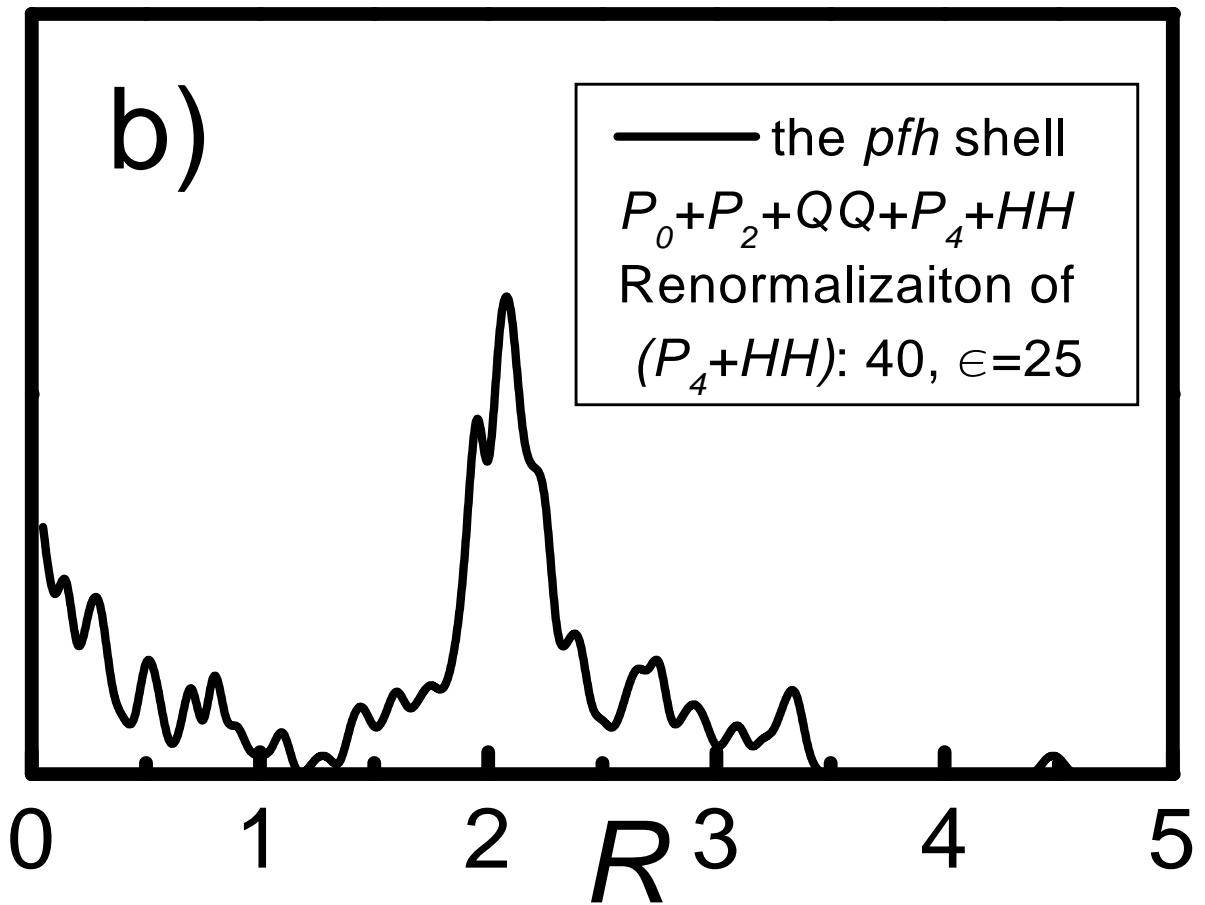
\end{document}